\documentclass{ws-p8-50x6-00}
\def\met{\mathbin{E\mkern - 11mu/_T}}
\begin{document}

\title{Experimental Studies of Diffractive Processes at the Tevatron}

\author{L. Demortier}

\address{Rockefeller University, 1230 York Avenue, New York, NY 10021-6399, USA\\
E-mail: luc@fnal.gov}

\author{(for the CDF and D\O\ Collaborations)}
\address{}


\maketitle

\abstracts{
We review the diffractive measurements made by the CDF and D\O\ 
collaborations during Run I at the Tevatron and summarize the
detector upgrades relevant for the Run II diffractive physics program.
}

\section{Introduction}
Run I at the Tevatron lasted from 1992 to 1996.  During that time, the
CDF and D\O\ experiments conducted an extensive program of diffractive
studies on soft single and double diffraction, several diffractively
produced hard processes, and double-Pomeron exchange.  After reviewing the 
results obtained by each experiment, we describe the forward detector upgrades 
for Run II.

\section{Run I diffractive physics results from CDF}
CDF measured the single diffraction differential cross section at 
$\sqrt{s}=546$ and 1800 GeV.\cite{abe94}  A fit of the data at both
energies to the standard triple-Pomeron Regge formula for single
diffraction dissociation shows that Regge theory correctly predicts
the dependence on the rapidity-gap width $\Delta\eta$ 
($\eta\equiv-\ln\tan\frac{\theta}{2}$) for 
$\Delta\eta>3$, which corresponds to a leading proton fractional
momentum loss $\xi<0.05$.  However, the $\sqrt{s}$ dependence of the
overall normalization is not correctly modeled, the data being 
an order of magnitude lower than the prediction at $\sqrt{s}=1800$ GeV.

The {\em double}-diffraction differential cross section was recently studied 
with a sample of minimum-bias events, i.e. events collected by triggering on 
a beam-beam counter coincidence between the proton and antiproton sides of 
the detector.\cite{affo01a}  The diffractive event rate is approximately flat 
in $\Delta\eta_{exp}^{0}\equiv\eta_{\max}-\eta_{\min}$, where 
$\eta_{\max}$ ($\eta_{\min}$) is the pseudorapidity of the particle closest to
$\eta=0$ in the (anti)proton direction.  In contrast, the non-diffractive
event rate decreases exponentially with increasing $\Delta\eta_{exp}^{0}$.
The data at both $\sqrt{s}=630$ and $1800$ GeV clearly exhibit a 
double-diffractive component for $\Delta\eta_{exp}^{0}>3$, and the 
$\Delta\eta_{exp}^{0}$ dependence of this component is
correctly described by Regge theory and factorization.
The measured double diffraction cross sections are:
\begin{eqnarray*}
\sigma_{DD}(\sqrt{s}=630\;{\rm GeV},\Delta\eta>3) & = & 
4.58\pm 0.02\,{\rm (stat)}\pm 1.5\,{\rm (syst)}\;{\rm mb},\\
\sigma_{DD}(\sqrt{s}=1800\;{\rm GeV},\Delta\eta>3) & = & 
6.32\pm 0.03\,{\rm (stat)}\pm 1.7\,{\rm (syst)}\;{\rm mb}.
\end{eqnarray*}
These measurements are about an order of magnitude lower than
expectations based on the triple-Pomeron amplitude.

Single diffractive events can be identified by the presence of a ``rapidity 
gap'', i.e. a rapidity interval devoid of particles, in the forward region on one
side of the detector.  Using this signature, CDF has measured the ratio $R$ of 
single-diffractive to non-diffractive rates of events containing a $W$-boson\cite{abe97b},
a dijet\cite{abe97a}, a $b$-quark\cite{affo00a}, and a $J/\psi$ meson\cite{affo01b} 
at $\sqrt{s}=1800$ GeV.  As shown in Table~\ref{Tab:ratios}, all these ratios
are of order 1\%.  Since the processes listed have different sensitivities to 
the quark and gluon content of the Pomeron, the ratios can be used to extract
the gluon fraction $f^{I\!\!P}_{g}$ in the Pomeron.
An analysis of the $W$, dijet, and $b$ results yields
$f^{I\!\!P}_{g}=0.54^{+0.16}_{-0.14}$.  The same analysis also shows that all the
diffractive to non-diffractive ratios are consistently suppressed by a factor
$D=0.19\pm 0.04$ with respect to expectations based on HERA measurements and QCD
factorization, indicating a breakdown of the latter.
\begin{table}
\caption{Ratio $R$ of diffractive to non-diffractive event rates for various
         processes.
         \label{Tab:ratios}}
\begin{center}
\begin{tabular}{|lccll|}
\hline
Process   &    $\sqrt{s}$ (GeV)  &  $R$ (\%)   &  \multicolumn{2}{c|}{Kinematic region} \\
\hline
$W(\rightarrow e\nu)$        & 1800 & $1.15\pm 0.55$ & $E_{T}^{e}>20$ GeV,  & $\met>20$ GeV      \\
Dijet             & 1800 & $0.75\pm 0.10$ & $E_{T}^{\rm jet}>20$ GeV, & $|\eta^{\rm jet}|>1.8$   \\
$b(\rightarrow eX)$          & 1800 & $0.62\pm 0.25$ & $E_{T}^{e}>9.5$ GeV, & $|\eta^{e}|<1.1$   \\
$J/\psi(\rightarrow\mu\mu)$  & 1800 & $1.45\pm 0.25$ & $p_{T}^{\mu}>2$ GeV, & $|\eta^{\mu}|<1.0$ \\
\hline
Jet-G-Jet         & 1800 & $1.13\pm 0.16$ & $E_{T}^{\rm jet}>20$ GeV, & $|\eta^{\rm jet}|>1.8$   \\
Jet-G-Jet         & 630 & $2.7\pm 0.9$ & $E_{T}^{\rm jet}>8$ GeV, & $|\eta^{\rm jet}|>1.8$   \\
\hline
\end{tabular}
\end{center}
\end{table}
Even though $J/\psi$ and $b$ production both occur mainly via initial-state gluon fusion,
their $R$ values disagree.  This discrepancy can be entirely ascribed to the difference
in kinematic region probed by the two processes.\cite{affo01b}

The bottom two lines of Table \ref{Tab:ratios} give the ratios of dijet event rates with
and without a gap between the jets.\cite{abe95,abe98a,abe98b}  
The measured fraction of color-singlet exchange is
constant as function of $\Delta\eta^{\rm jet}$, $E_{T}^{\rm jet}$ and $x_{bj}$.

A Roman pot spectrometer allows CDF to detect leading antiprotons and to determine 
$\xi$ and the four-momentum transfer squared $t$.  CDF measured the ratio of single-diffractive 
to non-diffractive dijet production rates at $\sqrt{s}=1800$\cite{affo00b} and 630 
GeV\cite{affo01c} as a function of the $x_{bj}$ of the struck parton in the $\bar{p}$.  
In leading order QCD, this ratio equals the ratio of the corresponding structure functions 
and can therefore be used to extract the diffractive structure function of the proton.
The CDF diffractive structure function is steeper than, and severely suppressed relative to
predictions based on extrapolations of the diffractive parton densities extracted by the
H1 collaboration from DIS measurements at HERA.

A search for exclusive dijet production by double-Pomeron exchange
($p+\bar{p}\rightarrow p^{\prime}+{\rm jet}_{1}+{\rm jet}_{2}+\bar{p}^{\prime}$)
yielded a 95\% C.L. upper
limit of 3.7 nb for $0.035<\xi_{\bar{p}}<0.095$ and jets with $E_{T}>7$ GeV and 
$-4.2<\eta<2.4$.\cite{affo00c}

\section{CDF forward detector upgrades}
\begin{figure}[hbt]
\epsfxsize=20pc
\epsfysize=12pc
\centerline{\epsfbox{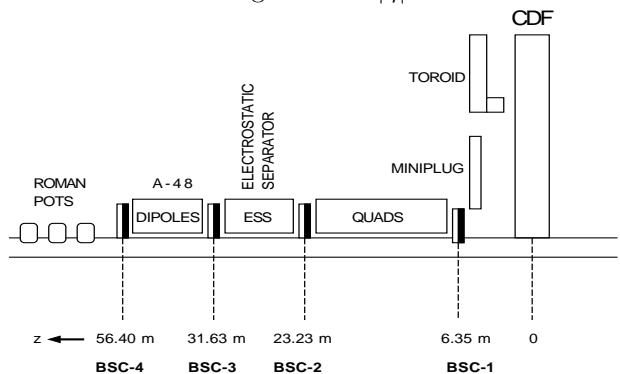}}
\vspace*{-2mm}\caption{Layout of the CDF forward detectors for Run II, on the antiproton side
of the central detector.  The proton side does not have a Roman pot spectrometer.
\label{fig:CDF_FD}}
\end{figure}

The CDF forward detector for Run II is shown schematically in Figure \ref{fig:CDF_FD}.
This system comprises a Roman Pot spectrometer to detect leading antiprotons and 
measure $\xi$ and $t$, beam shower counters on both sides of the central detector 
($5.5<|\eta|<7.5$) for triggering on events with forward rapidity gaps, and two 
``miniplug'' calorimeters in the region $3.5<|\eta|<5.5$.
\vspace*{-4mm}

\vspace*{-4.5mm}\section{Run I diffractive physics results from D\O\ }
The D\O\ collaboration extensively studied events with two jets separated by a rapidity gap
at $\sqrt{s}=1800$ and $630$ GeV.\cite{abac94,abac96,abbo98}  The measured color-singlet
fraction at $\sqrt{s}=1800$ GeV is $(0.94\pm 0.13)$\% and tends to increase with dijet 
$E_{T}$ and $\Delta\eta$.
This observation is consistent with a soft-color rearrangement model preferring initial
quark states.  Photon-like, U(1), and current two-gluon models do not adequately describe
the data.

More recently, D\O\  studied hard single diffraction at $\sqrt{s}=630$ and 1800 GeV.
\cite{abbo99}  At each energy, a sample of forward dijets (with the two leading jet 
pseudorapidities $\eta_{1}$, $\eta_{2}>1.6$ or $<-1.6$) and a sample 
of central dijets (with $|\eta_{1}|$, $|\eta_{2}|<1$) are defined.  Diffractive events are 
then identified 
by the rapidity-gap method.  Table~\ref{Tab:D0_ratios} shows the fractions of gap events 
for the four data samples and compares them with various models for the Pomeron structure 
function.
\begin{table}[ht]
\caption{Measured and predicted gap fractions and their ratios for the 630 and 1800 GeV
         forward (F) and central (C) dijet samples.  Predictions are given for four different 
         Pomeron structure functions $s(\beta)$: hard gluon ($s(\beta)\propto\beta (1-\beta)$), 
         flat gluon ($s(\beta)$ const.), soft gluon ($s(\beta)\propto (1-\beta)^{5}$) and quark 
         ($s(\beta)\propto\beta (1-\beta)$).
         \label{Tab:D0_ratios}}
\begin{center}
\begin{tabular}{|r@{$\;\;\;$}c@{$\;\;$}c@{$\;\;\;$}c@{$\;\;$}c@{$\;\;\;$}c|}
\hline
\multicolumn{6}{|c|}{Gap Fractions (\%)} \\[1mm]
Sample & Data & Hard G. & Flat G. & Soft G. & Quark \\
1800 F & $0.65\pm 0.04$ & $2.2\pm 0.3$ & $2.2\pm 0.3$ & $1.4\pm 0.2$   & $0.79\pm 0.12$ \\
1800 C & $0.22\pm 0.05$ & $2.5\pm 0.4$ & $3.5\pm 0.5$ & $0.05\pm 0.01$ & $0.49\pm 0.06$ \\
630 F 
       & $1.19\pm 0.08$ & $3.9\pm 0.9$ & $3.1\pm 0.8$ & $1.9\pm 0.4$   & $2.2\pm 0.5$   \\
630 C 
       & $0.90\pm 0.06$ & $5.2\pm 0.7$ & $6.3\pm 0.9$ & $0.14\pm 0.04$ & $1.6\pm 0.2$   \\
\hline
\multicolumn{6}{|c|}{Ratios of Gap Fractions} \\[1mm]
$\frac{630}{1800}$ F 
       &  $1.8\pm 0.2$  & $1.7\pm 0.4$ & $1.4\pm 0.3$ &  $1.4\pm 0.3$  & $2.7\pm 0.6$   \\[1.5mm]
$\frac{630}{1800}$ C 
       &  $4.1\pm 0.9$  & $2.1\pm 0.4$ & $1.8\pm 0.3$ &  $3.1\pm 1.1$  & $3.2\pm 0.5$   \\[1.5mm]
1800 $\frac{\rm F}{\rm C}$
       &  $3.0\pm 0.7$  &$0.88\pm 0.18$&$0.64\pm 0.12$&  $30.\pm 8.$   & $1.6\pm 0.3$   \\[1mm]
630 $\frac{\rm F}{\rm C}$
       &  $1.3\pm 0.1$  &$0.75\pm 0.16$&$0.48\pm 0.12$&  $13.\pm 4.$   & $1.4\pm 0.3$   \\[1mm]
\hline
\end{tabular}
\end{center}
\end{table}
The predictions for a hard-gluon structure are far higher than supported by the data.
A quark structure yields better agreement, but has been shown by CDF to predict an
excessive rate of diffractive $W$ bosons.  Ratios of gap fractions provide additional
information, as the Pomeron flux factor cancels in ratios evaluated at the same $\sqrt{s}$.
The bottom two lines of the table show that, in spite of this cancellation, a hard-gluon 
structure still does not model the data.  D\O\ concludes that a gluon-dominated Pomeron,
containing both soft and hard components, and combined with a flux factor that decreases
with increasing $\sqrt{s}$, could describe all the data samples.

\section{D\O\ forward detector upgrades}
\begin{figure}[htb]
\epsfxsize=28pc
\epsfysize=14pc
\centerline{\epsfbox{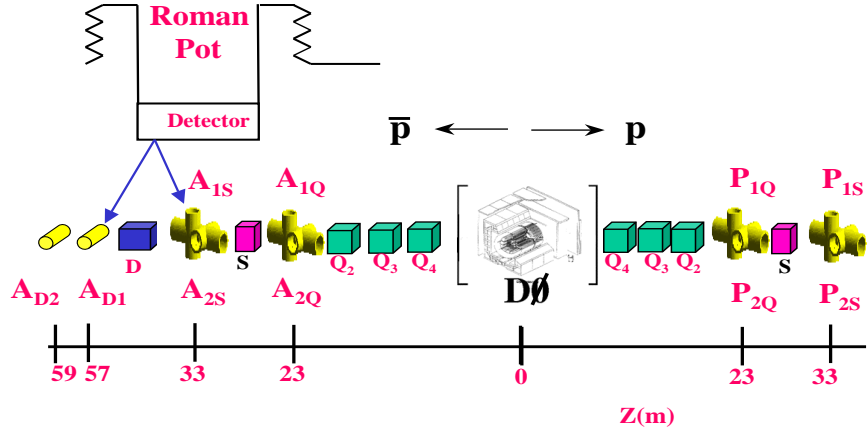}}
\vspace*{-3mm}\caption{Layout of the D\O\  forward proton detector, showing the central
detector (not to scale) at $Z=0$, the quadrupole magnets $Q_{i}$, dipole 
magnet $D$, and electrostatic separators $S$ from the accelerator lattice, 
and the roman pots housing the detectors for the quadrupole spectrometers $A_{1}$, 
$A_{2}$, $P_{1}$, and $P_{2}$ and for the dipole spectrometer $A_{D}$.
\label{fig:D0_fpd}}
\end{figure}

The Run II upgrade of the D\O\ detector includes a Forward Proton Detector
(FPD)\cite{bran97}, whose layout is shown in Figure \ref{fig:D0_fpd}.  The FPD 
consists of nine independent spectrometers, each of which uses magnets in the 
accelerator lattice together with position measurements along the (anti)proton 
track to determine the (anti)proton momentum and scattering angle.  There are eight 
quadrupole spectrometers, four on each 
side of the central detector, measuring tracks in both the horizontal and vertical 
planes, and one dipole spectrometer on the antiproton side.  This configuration 
has several advantages:
\begin{itemize}
\vspace*{-1mm}\item Having quadrupoles on both sides allows detection of elastic scattering
      events for alignment purposes and to monitor luminosity.
\vspace*{-1mm}\item Halo backgrounds can be rejected by diagonally opposite spectrometers,
      as in-time halo tracks on one side appear as early hits on the other.
\vspace*{-1mm}\item The possibility to tag both protons and antiprotons will allow unambiguous
      measurements of double-Pomeron exchange.
\vspace*{-1mm}\item A quadrupole spectrometer requires a minimum scattering angle, or $|t|$,
      to accept a proton, whereas a dipole spectrometer requires a minimum
      momentum loss $\xi$.  Therefore, the dipole spectrometer
      on the $\bar{p}$ side is crucial to calculate the acceptance of the
      quadrupole spectrometers, to avoid uncertainties in the extrapolation to
      $|t|=0$, and to provide large acceptance for high-mass events and double-Pomeron
      exchange.
\end{itemize}
\vspace*{-1mm}The spectrometers are incorporated into the standard D\O\ triggering and
data acquisition system, thereby enabling the collection of large samples of
diffractive events.  For 1 fb$^{-1}$ of integrated luminosity, D\O\ expects to
collect about 1K diffractive $W$ bosons, 3K hard double-Pomeron events, and
500K diffractive dijets.

\section{Diffractive physics program for Tevatron Run II}
Significant forward detector upgrades implemented by both D\O\ and CDF will
allow for a detailed study of the diffractive structure function of the proton,
including its process dependence, if any.  It should be possible to observe 
exclusive production of dijets by double Pomeron exchange, as well as 
low-mass exclusive states (glueballs?)  Jet-gap-jet events at high 
$\Delta\eta^{\rm jet}$ will provide a testing ground for BFKL.  Finally, there
is always the hope of making some unexpected discovery.

\section*{Acknowledgments}
I would like to thank Dino Goulianos and Andrew Brandt for their help with the 
preparation of this summary, and the symposium organizers and secretaries for
a very interesting and successful meeting.

\end{document}